\documentclass[twocolumn,showpacs,preprintnumbers,amsmath,amssymb]{revtex4}

\usepackage{graphicx}
\usepackage{dcolumn}
\usepackage{bm}
\usepackage{subfigure}
\usepackage{tabls}
\usepackage{caption}
\usepackage{booktabs,longtable}

\begin{document}


\title{ Self-affine Fractals Embedded in Spectra of Complex Networks\\}
\author{Huijie Yang$^{1,2}$}
\author{Chuanyang Yin$^{1,3}$}
\author{Guimei Zhu$^{1,3}$}
\author{Baowen Li$^{1,4}$}\email{phylibw@nus.edu.sg}%
\affiliation{$^1$ Department of Physics and Centre for
Computational Science and Engineering, National University of
Singapore,Singapore
117542  \\
$^2$ School of Management, University of Shanghai for Science and Technology, Shanghai 200093, China\\
$^3$ Department of Modern Physics, University of Science and
Technology of China, Hefei Anhui 230026, China\\
$^4$ NUS Graduate School for Integrative Sciences and Engineering,
Singapore 117597, Republic of Singapore}
\date{\today}

\begin{abstract}
The scaling properties of spectra of real world complex networks are
studied by using the wavelet transform. It is found that the spectra of
networks are multifractal. According to the values of the
long-range correlation exponent, the Hust exponent $H$, the networks can be classified into three types, namely, $H>0.5$, $H=0.5$ and $H<0.5$. All real
world networks considered belong to the class of $H \ge 0.5$, which may be
explained by the hierarchical properties.

\end{abstract}

\maketitle

Complex networks have attracted increasing attentions in recent
years due to their relevance to diverse problems in physical,
biological, and social sciences \cite{review,WattStrog98,Barab99}.
The primary purpose is to understand the relations between the
underlying structures, dynamics and functions. Generally, the
dynamical processes as the transport of mass, energy, signal and/or
information occur at differen structure scales. The organization
patterns at different scales may provide a reasonable solution to
the problems.

Song et al.\cite{Song05} found that the World-Wide-Web (WWW),
social, protein-protein interaction (PPI) and cellular networks are
fractal under a length-scale transform, namely, one can define a
topological box in which the shortest path between each pair is less
than $l_B$, the size of the box. The fractal behavior implies a
power-law relation between the minimum number of boxes, $N_B$,
needed to cover the entire network and the box size, $N_B(l_B) \sim
{l_B}^{-d_B}$. $d_B$ is the fractal dimension.

Detailed works have been done on the coverage methods \cite{Song07}.
It is shown that finding the minimum number of boxes to cover
networks can be mapped to the graph coloring problem in the
NP-complete complexity class, and the well-established algorithms in
the coloring problem provide a solution close to optimal. A random
burning-based algorithm is also proposed due to a number of other
benefits \cite{Goh06}.

A network with $N$ identical nodes is described by an adjacent
matrix $A$ whose elements $A_{ij}=1$ or $0$ if the nodes $i$
and $j$ are connected and disconnected, respectively. By mapping the
nodes and the edges to atoms and bonds, the network can be regarded
as a large cluster \cite{Yang04}. The Huckel Hamiltonian of the
large cluster reads, $\epsilon \cdot I+ \eta \cdot A$, where
$\epsilon$ and $\eta$ are the site energy and the hopping integral,
respectively. Generally, we can set $\epsilon=0$ and $\eta=1$, that
is, the Hamiltonian is $A$. The spectrum of the network is defined
as the rank ordered eigenvalues of $A$, namely, $E=\left\{E_1 \le
E_2 \le \cdots \le E_N \right\}$.

The topological structure of the network determines the spectrum.
The invariance properties embedded in the spectrum in turn reflect
the topological symmetries of the network. It is well known that the
fractal structures of aperiodic crystals lead to the fractal
behaviors of the corresponding spectra (For a detailed review, see
Ref. \cite{Tang87} and the references therein). An interesting
question is then, how the fractal structures of networks affect the
corresponding spectra. In this paper, we shall detect the scaling
properties embedded in spectra of networks.

The wavelet transform (WT) \cite{Ivanov99} is used to detect the
scaling properties. We consider the nearest neighbor level spacing
series $L=\left\{ L_i = E_{i+1}-E_{i}, i=1,2,\cdots,N-1 \right \}$.
The WT of the series $L$ can be calculated as, $T(s, a) =
\frac{1}{a}\sum\nolimits_{i = 1}^{N - 1} {L_i \cdot g\left( {\frac{i
- s}{a}} \right)}$. $g$ is the wavelet, $a$ the given scale. The
wavelet transform can remove effectively polynomial trends along the
series.

The series under consideration can be decomposed into many subsets
characterized by different local Hurst exponents, which quantify the
local singular behavior and thus relate to the local scaling of the
series. Traditionally, the local Hurst exponents are evaluated
through the modulus of the maximal values of $T(s, a)$ at each point
in the series. We denote the positions of the WT maximum with
$\left\{ s_1 ,s_2 , \cdots s_M \right\}$. In the long scale limit,
the partition function is expressed as,

\begin{equation}
\label{eq2} Z(a,q) = \sum\limits_{s = s_1 }^{s_M } {\left| {T (s,a)}
\right|} ^q\sim a^{\tau (q)}.
\end{equation}

\noindent For positive and negative $q$, $\tau (q)$ reflects the
scalings of the large and small fluctuations, respectively.

If $\tau(q)$ is a straight line, the analyzed series contains only
linear correlations (monofractal) and its slope represents the Hurst
exponent. If $\tau(q)$ is a nonlinear function, the series is called
multifractal, since different subsets of the series exhibit
different local Hurst exponents. In order to characterize this
multifractal, one considers the fractal dimensions of the subsets of
the series that is characterized by $\alpha(q)$, which is related to
$\tau(q)$ by a Legendre transorm, $D(h) = qh - \tau (q), h =
\frac{d\tau (q)}{dq}$. The width of this function for
$q\rightarrow\pm\infty$ is a measure for the strength of
multifractal, $\Delta \alpha=\alpha_{max}-\alpha_{min}$.

However, the numerical derivative of $\tau(q)$ in this method may
induce unacceptable errors to $\Delta \alpha$. Thus, we employ a
functional form fitted to $\tau(q)$ suggested by Kantelhardt et al.
\cite{Kantelhardt05},

\begin{equation} \label{eq4} \tau (q) = -
\ln \left( {x^q +y^q} \right)/\ln 2.
\end{equation}
\noindent The distribution width of the Hurst exponent is given by,
\begin{equation}
\label{eq3} \Delta \alpha = \left| {\ln x - \ln y} \right|/\ln 2.
\end{equation}

\noindent Sometimes the bifractal is required to obtain $\Delta
\alpha$. For a bifractal series $\tau(q)$ is characterized by two
distinct slopes $\alpha_1$ and $\alpha_2$,
\begin{equation}
\label{eq3} \tau (q) = \left\{ {{\begin{array}{*{20}c}
 {q\alpha _1 - 1} \hfill & {q \le q_x } \hfill \\
 {q\alpha _2 + q_x \left( {\alpha _1 - \alpha _2 } \right) - 1} \hfill & {q
> q_x } \hfill \\
\end{array} }} \right.
\end{equation}

\noindent or

\begin{equation}
\label{eq3}\tau (q) = \left\{ {{\begin{array}{*{20}c}
 {q\alpha _1 + q_x (\alpha _2 - \alpha _1 ) - 1} \hfill & {q \le q_x }
\hfill \\
 {q\alpha _2 - 1} \hfill & {q > q_x } \hfill \\
\end{array} }} \right..
\end{equation}
\noindent We can obtain the multifractal strength,
$\Delta\alpha=\left| \alpha_1-\alpha_2\right|$. These forms can be
derived from a modification of the multiplicative cascade model
\cite{Koscielny06}.

In the multifractal case, one conventionally refers to the second
moment as Hurst exponent, i.e.,

\begin{equation}
\label{eq3} H=(\tau(2)+1)/2.
\end{equation}

\noindent For $H>0.5$, the levels will tend to form local clusters
with small level spacings in different scales, while for $H<0.5$
these clusters can not be formed. The critical value of $H=0.5$
corresponds to a series that the corresponding integrated series
behaves like a random walk. These characteristics are induced
obviously by the structures of networks generated by different
mechanisms. Therefore, the exponent $H$ can be used as a criterion
to classify networks into three categories with $H< 0.5$, $H=0.5$
and $H> 0.5$, respectively.

Theoretically, we should have $\tau(0)=-1$ while the calculated
values may deviate slightly from it. The deviation
$\Delta\tau_0=1-\left|\tau(0)\right|$ can be used as the estimation
of the error of $\tau(2)$. The corresponding error of $H$ is,

\begin{equation}
\label{eq3} \delta H=\Delta\tau(0)/2.
\end{equation}

In each application reported below we have used the real analytic
wavelet $g^{(n)}$ among the class of derivatives of the Gaussian
function. The polynomial trends up to $n$ order can be removed. We
present the results by using the parameter value $n=4$. Calculations
with higher orders ($n=5$ and $6$) lead to almost the same results.

Randomizing $L$, we detect also the scaling behaviors embedded in
the resulting series (called shuffled series) as a comparison. The
partition function $Z(a,q)$ are calculated by using the software
provided in PhysioTookit \cite{Goldenber00}. The integrated series,
i.e., the spectrum $E$ is used as the input data. The relation in
$Eq.(6)$ is also checked by using the DFA software.

\begin{figure}
\scalebox{0.2}[0.2]{\includegraphics{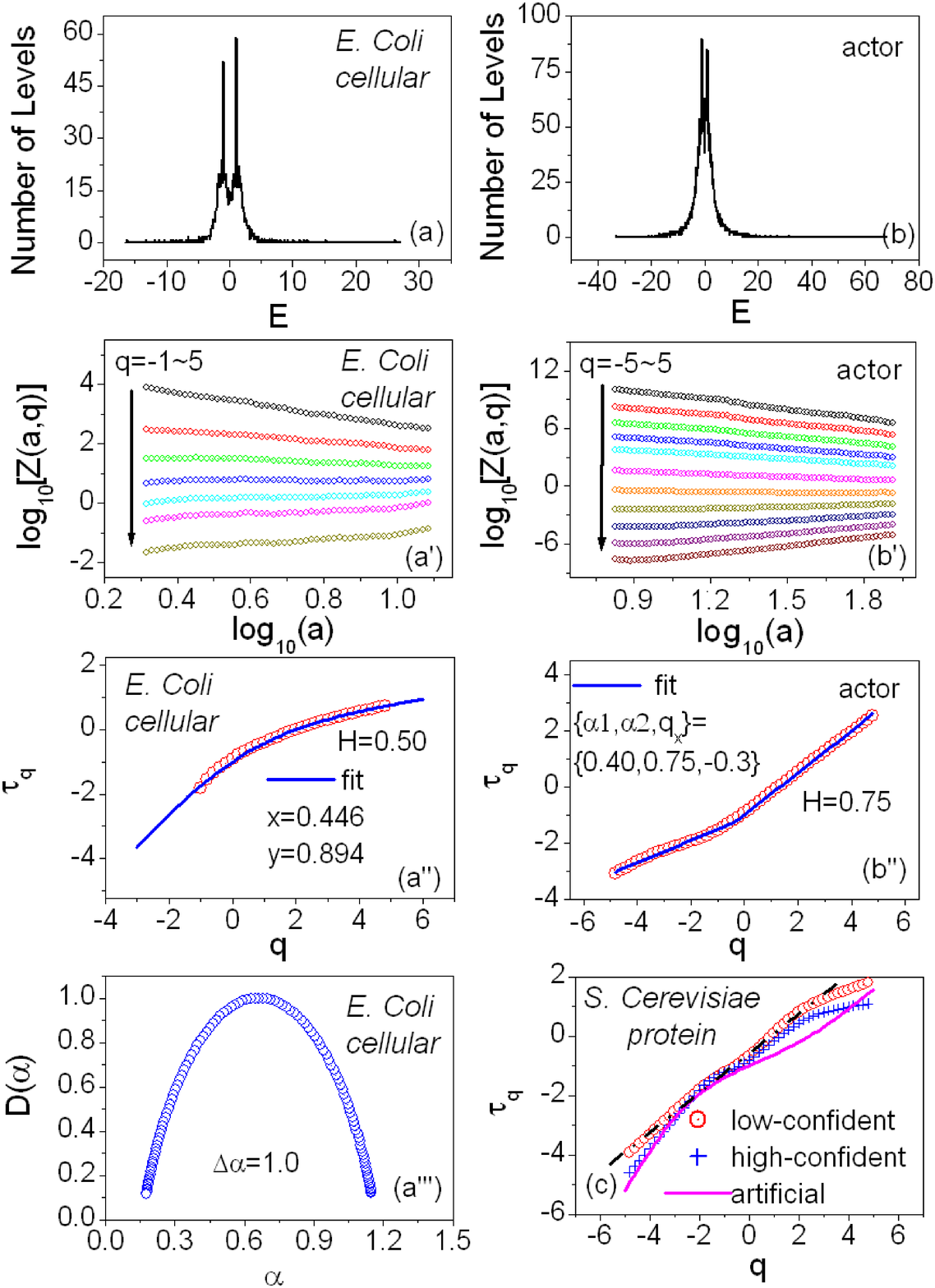}}
\caption{\label{fig:epsart} (Color online) Self-affine properties
embedded in spectra of real world networks. (a,b) histograms of the
levels of the E. Coli cellular network and the actor subnetwork
(containing the nodes numbered 1-8,000). (a',b') partition
functions. (a'',b'') scaling exponent $\tau_q$ as a function of $q$.
The E. Coli cellular network behaves multifractal. The actor
subnetwork behaves bifractal, and $Eq.(5)$ is used to obtain $\Delta
\alpha$. (c) The relations of $\tau_q$ versus $q$ for the
high-confident, low-confident and artificial versions of the S.
cerevisiae protein-protein interaction network. High-confidence may
not necessarily imply high quality. The result for the artificial
network is an average over 20 realizations. A dashed line is added
as reference, the slope of which is 0.66. The partition functions
are shifted to avoid overlapping.}
\end{figure}

We examine the scaling behaviors for the spectra of some real world
networks \cite{Song05}. The cellular networks consider the cellular
functions as intermediate metabolism and bioenergetics, information
pathways, electron transport, and transmembrane transport. The
direct edges are replaced simply with non-directed edges. Generally,
we find the spectra of real world networks to be multifractal. We
present in Fig.1(a)-(a''') the result for the E. Coli cellular
network. The Hurst exponent distributes in a wide range, $\Delta
\alpha = 1.0$. The long-range correlation exponent is $0.5$. For the
actor network, we consider only the subnetwork containing the nodes
numbered 1-8,000. The spectrum for this network behaves bifractal,
and the long-range correlation exponent is 0.75, as given in
Fig.1(b)-(b'').

For the S. cerevisiae protein-protein interaction network, we
consider two versions of the database. One is investigated by
 \cite{Song05}, which contains 1381 nodes and 2493 edges. The other
one is from \cite{Han04}, which has 1037 nodes and 1058 edges. The
edges in this version are high-confident. They are called
low-confident and high-confident networks, respectively. As shown in
Fig.1(c), the addition of the so-called low-confident edges in the
low-confident network makes $\tau(q)$ versus $q$ significantly
closer to a linear relation. The slope of the black dashed line is
$H=0.66$.

This change of the relations of $\tau_q$ versus $q$ for the
high-confident and low-confident networks may be caused simply by a
size-effect. To exclude the size-effect, we consider also some
artificial networks, in which the same number of random edges and
nodes are added to the high-confident network. Starting from the
high-confident network, at each step a new node is added by
connecting it with a randomly selected node in the existing network.
When the size of the network reaches 1381, we add edges between
randomly selected pairs of nodes until the total number of edges is
2493. The resulting network has the same numbers of edges and nodes
with the low-confident network.

The randomly added edges and nodes in the artificial networks do not
lead to similar result. This comparison may prefer to support the
conjecture that an exactly constructed protein interaction network
behaves perfectly fractal. The deviation of the actual structure
from the perfect fractal is due to the incompleteness of the
databases which are continuously being updated with newly discovered
physical interactions. The high-confidence may not necessarily imply
high-quality.

The scaling characteristics for the real world networks are listed
in Table I. We present only the results of networks whose partition
functions meet the scaling relation in $Eq.(1)$ in a wide range of
$q$, namely $\Delta q \ge 5$. Interestingly, we find that the values
of $H$ for the real world networks are in the range of $H \ge 0.41$.
Taking note of the values of error estimations $\delta H$, as
presented in Table I, we have $H \ge 0.41 \approx 0.5$.

The hierarchical property may be helpful in understanding the fact
that $H \ge 0.5$ for the real world networks. In the present paper,
we use the definition of hierarchy proposed in \cite{Ravasz03}. That
is, for a hierarchical network, besides the small-world and
scale-free characteristics, there exists a simple relation between
the clustering coefficient $C$ and the degree $k$, $C(k) \sim
k^{-1}$. Our detailed calculations show that all the considered real
world networks are hierarchical in this sense.

For the Watts-Strogatz small-world (WSSW) networks
\cite{WattStrog98}, we can construct a regular circle lattice, with
each node connected with its $d$ right-handed nearest neighbors.
Each edge is rewired with probability $p_r$ to another randomly
selected node.

As for the Barabasi-Albert scale-free (BASF) networks
\cite{Barab99}, we start from several connected nodes as a seed, at
each step we add a new node and $w$ edges from the new node to
different preferentially selected nodes in the existing network. The
probability for a node being selected is proportional to its degree.

The results for the constructed networks are listed in Table I. The
sizes of the networks are $2,000$. And the parameter $d$ is assigned
$2$. The values of $H$ for the WSSW networks are in the range of
$0.15 \sim 0.31$. For the BASF networks, with the increase of $w$
the small-world effect becomes more and more significant and the
value of $H$ decreases rapidly from $0.5$ ($w=2$) to $< 0.2$
($w>3$). Hence, for the real-world networks, the hierarchy is
essential for the values of $H$ being larger than $0.5$.

The values of $H$ for the shuffled series are almost exactly $0.5$.
And the corrections due to the fluctuations $\Delta \tau(0)$ are
neglectable. The WSSW and BASF networks with sizes $4,000,6,000$ and
$8,000$ have similar characteristics (not shown in Table I).

In summary, we have found self-affine fractals embedded in spectra
of complex networks. For the real world networks considered in the
present work, the values of the long-range correlation exponents are
in the range of $H \ge 0.5$, which may be attributed to the
hierarchical properties in the sense of a dependence of clustering
on the degree. This evidence may support the idea that fractals in
topological structures induce the fractals in spectra of networks.

For the constructed BASF networks, which have not box-based fractal
structures, we have also found rich multifractal structures in the
spectra. However, the values of $H$ are all significantly smaller
than $0.5$ for networks with $w\ge 3$. There may exist a new kind of
scale-invariance in the topological structures rather than the
box-based fractals in the constructed networks.

One paradox may be raised that the box-based fractal can be
explained with degree-degree anti-correlations \cite{Song05}, while
we find the positive correlations in the spectra ($H>0.5$) for the
real world networks. Because of the degree-degree anti-correlations,
the nodes tend to aggregate into many small-sized structure clusters
with the hubs as centers. And there exist loosely connections
between the clusters. There are strong ''repulsive effects" between
the levels within each cluster, but the levels for different
clusters may be very close or even degenerate. That is, there will
appear some locals with high density of levels in the spectrum,
called level clusters. We can expect $H>0.5$ (positive correlations
in spectra) for this kind of networks.

While for the BASF networks with $w\ge 3$, since the strong
correlations between the hubs, the clusters centered at the hubs
will merge into a small number of large-sized clusters. The strong
"repulsive effects" between the levels make the so-called
"clustering of levels" impossible. Consequently, the spectra are
anti-correlated ($H<0.5$).

Hence, the difference of our results with the box-based results is
not necessarily a contradiction. Obviously, the relation between the
self-affine behaviors of spectra and the fractal dimension based
upon box-counting approaches deserves further investigation.

Network comparison is an important topic in systems biology. It can
shed light on the evolutionary and diseases detecting by comparing
cellular networks of different species or diseased and healthy
cellular networks \cite{James07}. One basic task is to design node
labeling-independent representations of networks and circumvent the
problem of graph isomorphism. Spectra analysis of complex networks may
provide useful information for that purpose.

The work is supported by the NUS Faculty Research Grant No.
R-144-000-165-112/101. It is also supported in part by the National
Science Foundation of China under Grant Nos.70571074 and 10635040. We are very grateful to one of the anonymous referee for very detailed suggestive and helpful comments.

\begin{longtable*}[htbp]
{|p{45pt}|p{75pt}|p{68pt}|p{50pt}|p{0.1pt}|p{35pt}|p{65pt}|p{68pt}|p{50pt}|}
\hline \multicolumn{2}{|p{122pt}|}{Networks} & $x / y / \Delta
\alpha $ \par or \par $\left\{ {\alpha _1 ,\alpha _2 ,q_x }
\right\}$& $H / \delta H$& \raisebox{-40.50ex}[0cm][0cm]{}&
\multicolumn{2}{|p{114pt}|}{Networks} & $x / y / \Delta \alpha $
\par or \par $\left\{ {\alpha _1 ,\alpha _2 ,q_x } \right\}$&
$H / \delta H$ \\
\cline{1-4} \cline{6-9} \multicolumn{2}{|p{122pt}|}{WWW[3] } &
{\{}0.00,0.79, 0.00{\}}& 0.97/0.08&
 &
\raisebox{-10.50ex}[0cm][0cm]{WSSW}&$p_r = 0.00$& 0.50/0.50/0.00&
1.02/0.01 \\
\cline{1-4} \cline{7-9} \multicolumn{2}{|p{113pt}|}{Actor[3]} &
{\{}0.40,0.75,-0.30{\}} & 0.75/0.03&
 &
 &
$p_r = 0.03$& 0.81/0.81/0.00&
0.31/0.01 \\
\cline{1-4} \cline{7-9} \raisebox{-1.50ex}[0cm][0cm]{PPI [12]}&
\textit{D .melanogaster}& {\{}0.51,0.66, 0.40{\}}& 0.66/0.01&
 &
 &
$p_r = 0.12$& 0.76/0.95/0.33&
0.22/0.01 \\
\cline{2-4} \cline{7-9}
 &
\textit{C. elegans}& 0.41/0.67/0.73& 0.85/0.07&
 &
 &
$p_r = 0.15$& 0.66/1.00/0.61&
0.24/0.01 \\
\cline{1-4} \cline{7-9} \raisebox{-15.00ex}[0cm][0cm]{Cellular [3]}&
\textit{B. burgdorferi}& 0.61/0.61/0.00& 0.72/0.07&
 &
 &
$p_r = 0.21$& 0.76/1.00/0.40&
0.17/0.01 \\
\cline{2-4} \cline{7-9}
 &
\textit{A. aeolicus}& 0.42/0.80/0.94& 0.64/0.01&
 &
 &
$p_r = 0.24$& 0.70/1.00/0.50&
0.21/0.01 \\
\cline{2-4} \cline{7-9}
 &
\textit{C. elegans}& 0.37/0.93/1.35& 0.50/0.03&
 &
 &
$p_r = 0.27$& 0.72/1.00/0.47&
0.20/0.01 \\
\cline{2-4} \cline{7-9}
 &
\textit{E. coli}& 0.45/0.89/1.00& 0.50/0.04&
 &
 &
$p_r = 0.30$& 0.73/1.00/0.46&
0.19/0.02 \\
\cline{2-4} \cline{6-9}
 &
\textit{H. pylori}& 0.45/0.83/0.89& 0.59/0.08&
 &
\raisebox{-9.00ex}[0cm][0cm]{BASF}& w = 2& 0.71/0.71/0.00&
0.50/0.02 \\
\cline{2-4} \cline{7-9}
 &
\textit{M. leprae}& 0.51/0.88/0.77& 0.47/0.04&
 &
 &
$w = 3$& {\{}1.05,0.25,-1.13{\}} &
0.25/0.01 \\
\cline{2-4} \cline{7-9}
 &
\textit{P. aeruginosa}& 0.43/0.93/1.12& 0.46/0.02&
 &
 &
$w = 4$& 0.77/1.00/0.39&
0.17/0.00 \\
\cline{2-4} \cline{7-9}
 &
\textit{S. typhi}& 0.42/0.96/1.18& 0.43/0.04&
 &
 &
$w = 5$& 0.76/1.00/0.40&
0.17/0.02 \\
\cline{2-4} \cline{7-9}
 &
\textit{T. pallidum}& 0.38/0.82/1.11& 0.65/0.07&
 &
 &
$w = 6$& 0.74/1.00/0.43&
0.18/0.01 \\
\cline{2-4} \cline{7-9}
 &
\textit{Y. pestis}& 0.54/0.87/0.69& 0.47/0.03&
 &
 &
$w = 7$& 0.74/1.00/0.44&
0.19/0.01 \\
\cline{2-4} \cline{7-9}
 &
\textit{C. pneumoniae}& 0.63/0.86/0.44& 0.41/0.07&
 &
 &
$w = 8$& 0.79/1.00/0.35&
0.15/0.01 \\
\hline \caption{The self-affine fractals embedded in spectra of real
world, WSSW and BASF networks. For the real world networks the
values of $H$ are basically in the range of $H \ge 0.45\approx 0.5$,
while that for WSSW and BASF networks are significantly smaller,
namely, $H \le 0.3$. We present only the results for networks whose
partition functions meet the scaling relation in $Eq.(2)$ in a wide
range of $q$, namely $\Delta q \ge 5$.\\}
\end{longtable*}

\end{document}